# Photonic Cavity Synchronization of Nanomechanical Oscillators

Mahmood Bagheri[1†], Menno Poot[1†], Linran Fan[1], Florian Marquardt[2,3], Hong X. Tang[1*]

[1]Department of Electrical Engineering, Yale University, New Haven, CT 06511
[2]Institute for Theoretical Physics II, Universität Erlangen-Nürnberg, Staudtstr. 7, 91058 Erlangen, Germany
[3]Max Planck Institute for the Science of Light, Günther-Scharowsky-Straße 1/Bau 24, D-91058 Erlangen, Germany

Synchronization in oscillatory systems is a frequent natural phenomenon and is becoming an important concept in modern physics. Nanomechanical resonators are ideal systems for studying synchronization due to their controllable oscillation properties and engineerable nonlinearities. Here we demonstrate synchronization of two nanomechanical oscillators via a photonic resonator, enabling optomechanical synchronization between mechanically isolated nanomechanical resonators. Optical backaction gives rise to both reactive and dissipative coupling of the mechanical resonators, leading to coherent oscillation and mutual locking of resonators with dynamics beyond the widely accepted phase oscillator (Kuramoto) model. Besides the phase difference between the oscillators, also their amplitudes are coupled, resulting in the emergence of sidebands around the synchronized carrier signal.

[†] These authors contribute equally to this manuscript.

[*]Correspondence and requests for materials should be addressed to H. X. Tang. (e-mail: hong.tang@yale.edu).

Synchronization is a ubiquitous phenomenon where the phase difference between free-running oscillators remains constant due to the mutual coupling. Besides its well-accepted importance in biological sciences, today synchronization is becoming a powerful tool for many engineered systems [1]. For instance, synchronization is desirable in situations where high oscillating power, strong coherence, or low phase noise are needed, such as lasers [2], phase-locked loops [3], Josephson junction arrays [4], and spin-torque resonators [5]. Synchronization also promises to improve the accuracy of time-keeping devices [6]. Since the observations of synchronization in pendulums [7] this concept has found its bearings in science and engineering due to its potential applications in generating low-noise stable oscillating signals. Nanomechanical oscillators, on the other hand, are very appealing as they simultaneously offer high quality factor resonances, excellent scalability [8, 9] and are ideal systems for synchronization studies due to their highly engineerable nonlinearities [10].

However, achieving reproducible and strong coupling in nanomechanical devices remains difficult due to the unavoidable device non-uniformity and weak mutual coupling. This can be circumvented by coupling nanomechanical resonators to an optical cavity [11]. Recently, synchronization between two closely spaced micromechanical resonators was demonstrated using a hybrid optical mode of two coupled disk resonators and the synchronization phase-space was predicted using the Kuramoto model [12]. Here we experimentally demonstrate the first synchronization of two spatially separated nanoscale radio-frequency oscillators integrated inside an optical racetrack cavity. We show that this leads to a limit cycle in the reduced three-dimensional mechanical phase space (The two mechanical resonators' amplitudes and their phase difference [13]) and that the dynamics of two mechanical modes coupled via a common optical mode cannot be captured by the standard Kuramoto model [14]: as a result of the additional degrees of freedom of the coupled system, slow dynamics appear on top of the limit cycle, and sidebands emerge. These sidebands are true signatures of synchronized motion in the mechanical domain and are not to be confused with simple nonlinear intermodulation oscillatory modes. Their presence is important for the phase noise performance of synchronized optomechanical oscillators, and could counteract the common perception that synchronized states should always improve phase noise performance.

We investigate the interaction between two MHz-frequency nanomechanical resonators that are linked in an optical racetrack (Fig. 1(a)); The resonators are mechanically isolated, due to their large separation (~ 80 μm), ensuring that any coupling between them is through the optical field. The fabrication of these integrated photonic devices is readily scalable [13, 15], making this an ideal platform for synchronization studies [11,14]. The silicon beams are slightly buckled [15] and they may end up in the up or down state (Fig. 1(a)).

The measurement setup shown in Fig. 1d consists of a strong pump laser to create cavity backaction and a weak probe to detect the motion [13]. First the two nanomechanical resonators are characterized with the pump laser off, so that the backaction is small. When the two resonators are both in the buckled-up state their resonance frequencies are close with a small difference due to fabrication imperfections (6.53 MHz vs. 6.61 MHz) as indicated by the thermal displacement noise spectrum in Fig. 1(b). However, when one resonator is displaced from the up state to the down state, its mechanical resonance frequency drops to 4.05 MHz (Fig. 1(c)) due to asymmetries of the double-well potentials. The intrinsic damping rate of the resonators $\gamma/2\pi \sim 2$ kHz [13].

Optomechanical oscillators (OMOs) in single cavity, single resonator systems have been the subject of intense studies in recent years [16, 17, 18]. It is, however, expected that when multiple oscillators are embedded in a single cavity new phenomena will appear due to the mutual coupling via the cavity field. Figure 2a displays the evolution of the RF power spectrum of the transmitted probe laser light (the pump laser is blue detuned at $\Delta_0/\kappa = 0.3$, where $\Delta_0$ is the detuning between the laser and cavity frequency, and $\kappa$ is the cavity linewidth). At the lowest pump powers the thermomechanical motion of each resonator is visible as two lines at 4.0 and 6.5 MHz respectively. Upon increasing the pump power in this first regime, their Brownian motion is amplified as the backaction reduces the damping. Also, the optical spring effect [19] is visible as an increase in the resonance frequencies. Both effects are stronger in the 4 MHz resonator since in the buckled down state the effective refractive index of the racetrack depends more strongly on the displacement yielding a larger optomechanical coupling $g_{om}$ [15]; From simulations we estimate $g_{om}/2\pi = 140$ and 500 MHz/nm for up and down. The difference in backaction confirms that optical backaction is stronger than photothermal effects as the latter would be the same on both resonators, irrespective of their separation from substrate [20].

When the pump is increased beyond -2.7dBm, the optomechanical gain fully compensates the mechanical damping of the resonator with the lowest threshold, which we will label as "1", which starts to self-oscillate. This demarcates the onset of regime II which ranges from -2.7 dBm to 0.2 dBm. Yet, even though the oscillation amplitude of resonator 1 increased dramatically, the thermal motion of resonator 2 is undisturbed and is still clearly visible in Fig. 2(a) and (b). Note, that the frequency difference between the two resonators (2.5 MHz) is much larger than the spectral width of resonator 2, so the cavity occupation oscillation induced by OMO 1 cannot efficiently drive the other resonator. Regime II thus corresponds to the single oscillator system that we studied previously [15]. One might expect that resonator 2 simply starts to oscillate when the power is increased beyond the second threshold which is higher due to its lower $g_{om}$. Instead, different dynamics is encountered (Regime III) where the power spectrum displays a large magnitude over a wide range of frequencies. Here, the motion is chaotic and the phase, amplitude, and frequency of the transmitted probe light change on short timescales. This chaotic behavior only exists for a limited power range and vanishes beyond 0.7 dBm.

Above 0.7dBm the two oscillators start their synchronized motion, as evidenced by a dramatic change in the mechanical displacement spectrum. In Regime IV the detected photocurrent contains a strong narrow tone and, more importantly, the thermomechanical motion of resonator 2 is now no longer visible. The single strong peak in the RF spectrum indicates that both resonators are oscillating at the same frequency; In this case they are said to be synchronized since the phase difference between them remains constant. We have thus synchronized the two resonators despite the extremely large frequency difference: the second mode, originally at 6.7 MHZ, was almost twice as fast as the oscillations of the first mode at 3.9 MHz, indicating the extremely strong optomechanical interactions and the tunability of the double well potential in our system [15]. Note, that in previous micromechanical synchronization experiments the frequency difference was limited to ~0.2% [12].

The spectra in Regime IV also reveal another surprise: Sidebands emerge around the strongest peak (the carrier), indicating a modulation of the self-sustained oscillations. A close inspection of Fig. 2(a) and (e) shows that the spectra contain two equally spaced sidebands ~100-500 kHz from the carrier. Their presence implies a deteriorated signal phase noise at that particular sideband offset frequency [21]. These weak (~20-35 dBc), but clearly defined, sidebands are not transient phenomena as they persist during the entire data-acquisition time, which is much longer than the damping time of the resonator $\gamma^{-1}$. Also, in Regime II (Fig. 2(c)) with only a single OMO present, sidebands are absent, ruling out low-frequency thermal instabilities [22] interfering with optomechanical oscillations. Finally, the mixing of the sidebands in Regime IV with the strong carrier produces a signal below 500 kHz). This down-converted peak is not mechanical in origin but is due to the nonlinearity of the transduction at large amplitudes. Similarly, in Regime II the thermal motion of resonator 2 is mixed with that of OMO 1, resulting in a peak near 2.5 MHz.

To understand the origin of synchronization and the slow dynamics in the cavity-coupled oscillators we theoretically analyse this system; for details see [13]. Multiple uncoupled oscillators will each oscillate at their own frequency, but the cavity field couples the oscillators enabling synchronization as will be shown. When the frequency difference between the resonators $\upsilon \ll \bar{\Omega}$, the equations of motion for their complex amplitude $U_k = g_{om,k}\langle(u_k + \dot{u}_k/i\bar{\Omega})\exp(-i\bar{\Omega}t)\rangle = A_k \exp(i\theta_k)$ in the frame rotating at the average frequency $\bar{\Omega} = \frac{1}{2}(\Omega_1 + \Omega_2)$ become [14, 23,24]:

$$\dot{U}_{1,2} = \pm i\frac{\upsilon}{2}U_{1,2} - \frac{\gamma_{1,2}}{2}U_{1,2} - i\bar{\Omega}c_{om,1,2}\Phi(t), \qquad (1)$$

where $u_k$ are the displacements of the two resonators ($k$=1,2) and $\Phi(t)$ describes how the photon occupation responds to a dynamic displacement [24]. $c_{om,k} = \hbar n_{max} g_{om,k}^2/m_k\bar{\Omega}^3$ are the coupling strengths *and* $n_{max}$ is the maximum number of photons in the cavity. When multiple resonators are present, the cavity only feels the *combined* effect of all resonators. The same reasoning as for a single OMO shows that for multiple resonators coupled to the same cavity, the cavity response is $\Phi(t) = \Phi(A_+(t))$ where $A_+$ is the magnitude of the summed complex amplitude $U_+(t) = \sum_k U_k(t) = A_+(t)\exp(i\theta_+(t))$. $A_+$ depends on the phase difference between the individual oscillators, but not on the overall phase $\theta_+$. The equations of motion for the two OMOs are thus coupled together via $U_+$ and the cavity response function $\Phi(A_+)$. Synchronized motion of the two oscillators implies that they rotate at the same frequency $\bar{\Omega} + \epsilon$. Hence, $U_{1,2}(t) = Y_{1,2}\exp(i\epsilon t)$ must be a solution to Eq. (1): if no such solution exists, synchronization cannot take place. Inserting the solution into Eq. (1) yields the equation that determines the combined amplitudes of the limit cycles:

$$\Phi(A_+) = \frac{1}{\bar{\Omega}} \frac{\left(\frac{v}{2}\right)^2 - \epsilon^2 + i\epsilon\bar{\gamma} + \frac{i\,\delta\,\gamma v}{4} + \left(\frac{\bar{\gamma}}{2}\right)^2 - \left(\frac{\delta\gamma}{4}\right)^2}{2\epsilon\bar{c}_{om} + v\frac{\delta c_{om}}{2} + i\frac{\delta c_{om}}{2}\frac{\delta\gamma}{2} - i\bar{\gamma}\bar{c}_{om}}, \qquad (2)$$

where $\bar{c}_{om}$ ($\delta c_{om}$) and $\bar{\gamma}$ ($\delta\gamma$) are the average (difference in) coupling strengths and damping rates respectively. Solving Eq. (2) can be done as illustrated in Fig. 3(a); the left hand side is a curve in the complex plane parameterized by $A_+$ (for a given detuning and decay rate), whereas the right hand side depends only on the oscillators' properties and is parameterized by the unknown frequency $\epsilon$. Intersections of the two curves are thus solutions to Eq. (2). When inserting the obtained values for $A_+$ and $\epsilon$ back into Eq. (1) the individual contributions $Y_1$ and $Y_2$ can be obtained including their phases relative to the carrier $Y_+ = Y_1+Y_2$. Fig. 3(b) shows that for the parameters of Fig. 3(a) the two complex amplitudes $Y_1$ and $Y_2$ have a similar, but not identical, magnitude and oscillator 1 moves ahead of the second. Finally, for sufficiently asymmetric oscillators, the two curves can intersect more than once, leading to multistability [23,24], even in the unresolved sideband regime where a single oscillator always has a unique amplitude.

Equation (2) thus yields the fixed points with synchronization. However, to understand the dynamics around the corresponding limit cycle, Eq. (1) can be expanded for small excursions and the eigenvalues can be found. There are three independent degrees of freedom: the two oscillation amplitudes and the phase difference between them. The fourth degree of freedom, the overall phase $\theta_+$, is not fixed yielding a zero eigenvalue. Depending on the values of the other three eigenvalues different types of behavior are possible: returning back to the fixed point without oscillations (overdamped), returning with oscillations (underdamped), or the fixed point is unstable. In the second case, when any of the two oscillators is displaced, e.g. due to the ever-present stochastic thermal force or photon shot noise, it will return back to the fixed point in an oscillatory fashion, which shows up as sidebands in the frequency domain. In contrast, for a single oscillator, there is only one mechanical degree of freedom, the oscillation amplitude $A$ and the corresponding eigenvalue is always real. Any small deviation thus overdampedly returns to the limit cycle and no sidebands are generated [24].

The analytical model thus hints that the observed sidebands are due to the thermal force noise acting on the oscillators. To further analyze the synchronization dynamics, the full coupled equation of motion of the mechanical resonators and that of the cavity field (cf. Eqs. (S1) and (S2)) are integrated numerically (see Supplemental Material [13]) with mechanical nonlinearities included. We chose to simulate a conceptually clear situation with nearly identical resonators (5% frequency difference and identical $g_{om}$). The effect of (thermal) force noise is accounted for by kicking the resonators away from their steady state oscillations (see the Suppl. Mat. for a simulation with a stochastic force instead of kicks). Figure 3(c) shows the evolution of the light field power spectrum as a function of $n_{max}$ (i.e., the pump power). Similar to the experiment (Fig. 2), at low power two weak peaks are visible which also tune upwards with increasing power due to the optical spring effect. Around $n_{max} = 30$ oscillations start, but now the two resonators immediately oscillate simultaneously [25]. As expected from our analytical theory of synchronized motion, sidebands appear in the spectrum, but only when force noise (i.e. kicks) is included. As illustrated in [13] the oscillators are truly phase locked in this regime, indicating full synchronization. When further increasing $n_{max}$ the oscillations grow towards the top of the potential barriers and hence the frequency goes down. When the barrier is crossed at $n_{max} = 134$ the detuning suddenly changes dramatically and the oscillations stop. However, they reappear at higher powers and above the barrier, just as in the experiment, the oscillation frequency increases with increasing power (see also Fig. 2a). Also note that the sidebands start out far from the carrier at the onset ($n_{max} = 335$) of the oscillations in this regime and that, just as in Fig. 2a, they converge towards the carrier with increasing power. Also, the stochastic simulation [13] shows sideband strengths of the same magnitude as observed experimentally, confirming their thermal origin. Finally, the simulations also reproduce bands with chaotic behavior with broad spectrums similar to the one in Fig. 2(d). The simulations thus qualitatively reproduce most of the features observed in the experiment, including the correct tuning behavior of the resonance frequency, the appearance of the sidebands of the synchronized resonators due to thermal force noise with the correct strength, and chaos. It is expected that even better agreement could be reached if the exact shapes of the mechanical potentials were known, and by including displacement-dependent optomechanical coupling coefficients, although we have to leave this to future work.

We have also studied the dynamics of a single mechanical oscillator in the presence of an external oscillator encoded in the light field. This is extremely important in the context of synchronizing a remote oscillator to an external clock and also further validates our model for optomechanical synchronization. To this end, the pump power is set between the oscillation thresholds of the first and second resonator (so that the latter does not play a role) and is modulated at frequency $\Omega_0 = 6.800$ MHz. When the modulation index (*m*) is zero, resonator 1 oscillates freely in the up-state at 6.804 MHz as shown in the bottom spectrum in Fig. 4. However, when the modulation is switched on the oscillations jump to $\Omega_0$, synchronizing the OMO to the external clock. Interestingly, sidebands appear again. A prominent feature that the location of the sidebands is not constant: the offset frequency increases with *m*. All of this is reproduced in the numerical simulations [13] showing that many of the phenomena observed in the two-OMO case can also understood in the conceptually simpler injection locking experiments [26].

Our technique of coupling mechanical oscillators via a single photonic bus creates a whole new platform for nonlinear studies. It will enable synchronization of large arrays of individual optomechanical elements with interesting new collective phenomena [27] and allows synchronization over arbitrarily long distances. Finally, by exploiting the memory storage capabilities of the double well resonators [15] we envision combining the information of the mechanical bits with synchronization. This could, for example, be used to perform conditional coupling of oscillators, an interesting future direction enabled by our cavity field coupling.

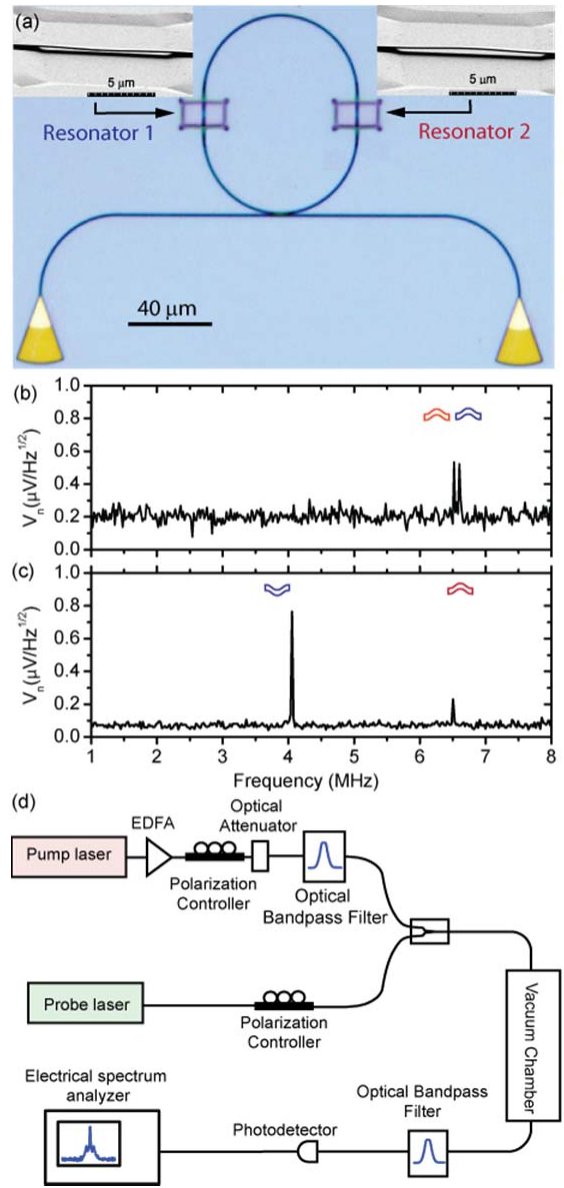

Figure 1. (a) Micrograph of a racetrack cavity with two 110nm x 500nm x 10um suspended portions as nanomechanical resonators. Insets show scanning electron micrographs of the mechanical resonators in buckled down (left) and buckled up (right) state. (b,c) thermal noise spectra in the up-up (b) down-up state (c). (d) The measurement setup with a weak probe laser, and a pump.

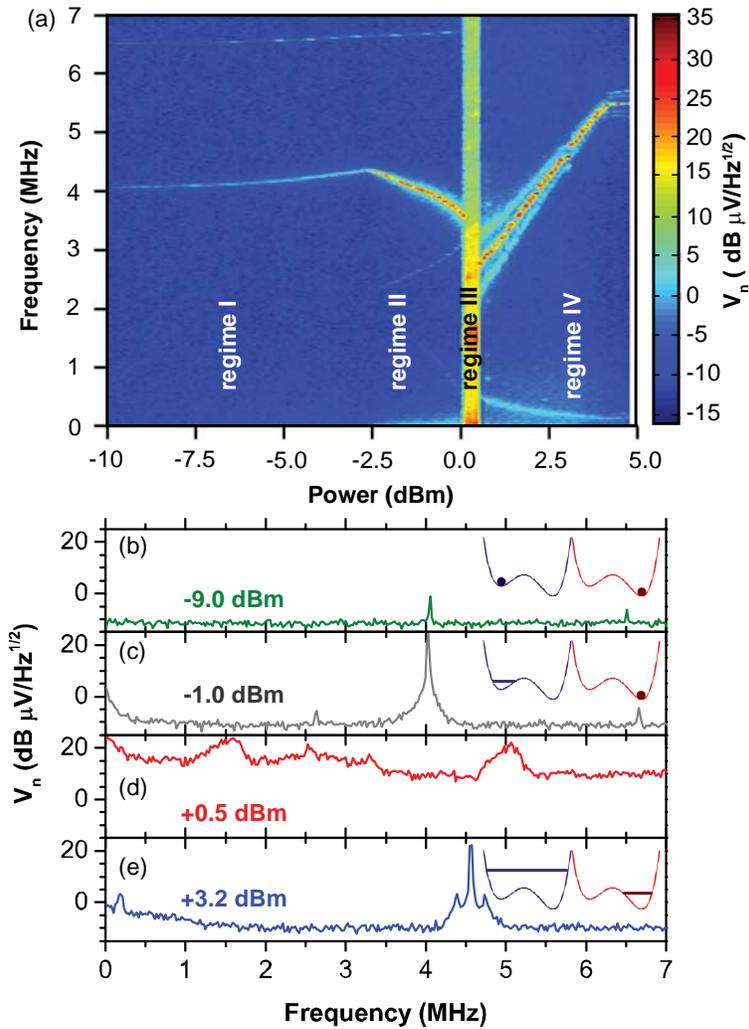

Figure 2. (a) The evolution of the RF power spectrum of the transmitted light as the pump power increases. (b)-(e) Cuts through panel (a) at the indicated pump power when both resonators are in a thermal state (regime I, (b)), (c) one resonator is in thermal motion while the other resonator experiences regenerative oscillations (regime II), (d) the chaotic regime (III) and (e) the two resonators are synchronized (regime IV). The insets schematically show the energy of resonators 1 (left) and 2 (right); dots correspond to small thermal motion, and lines to large oscillations.

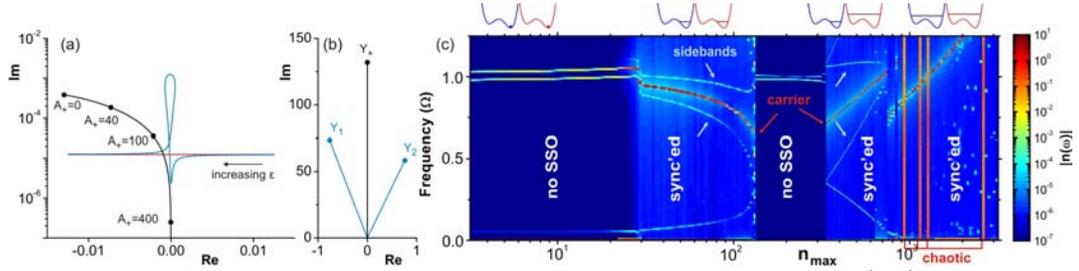

Figure 3. (a) Complex plane representation of the cavity response $\Phi(A_+)$ (black) and the right hand side of Eq. (5) for two identical oscillators (red) and two oscillators with $\bar{\gamma} = 0.001\ \bar{\Omega}$, $\bar{c}_{om} = 20$, $\upsilon = 0.01\ \bar{\Omega}$, $\delta\gamma = 0.0018\ \bar{\Omega}$, and $\delta c_{om} = 0$ (blue). The curves intersect at $A_+ = 132.0$ and $\epsilon = 0.0438\ \bar{\Omega}$ (b) Complex amplitudes of the individual oscillators for the solution shown in (a). (c) Colorplot of the Fourier transform of the photon number on a logarithmic scale. The carrier tone and its sidebands are indicated by red and white arrows respectively. The parameters are: $Q_1 = Q_2 = 6000$, $\kappa = 526\ \bar{\Omega}$, $\Delta_0 = 0.493\kappa$, $g_{om,1} = g_{om,2} = 1$.

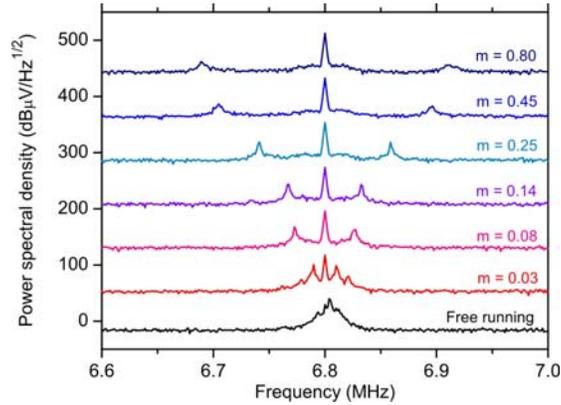

Figure 4. Measured RF power spectral density of the detector output with a free running oscillator (black) and oscillations in the presence of an increasingly larger modulation depth of the pump (red to dark blue) for constant *average* power. The curves are offset for clarity.

**Acknowledgements** We acknowledge funding support from a STIR grant from ARO and DARPA/MTO's ORCHID program through a grant from AFOSR. M.P acknowledges a Rubicon fellowship from the Netherlands Organization for Scientific Research (NWO) / Marie Curie Cofund Action. F.M. acknowledges an ERC Starting Grant, ITN cQOM, and a DFG Emmy-Noether grant. H.X.T. acknowledges support from a Packard Fellowship in Science and Engineering and a career award from the National Science Foundation. The authors thank Dr. M. Rooks of the Yale Institute for Nanoscience and Quantum Engineering for helping with the e-beam lithography, and Michael Power for helping with device fabrication.

# Supplementary Material for Photonic Cavity Synchronization of Remote Nanomechanical Oscillators

Mahmood Bagheri[1†], Menno Poot[1†], Linran Fan[1], Florian Marquardt[2,3], Hong X. Tang[1*]

## Contents

[†] These authors contribute equally to this manuscript.

*Correspondence and requests for materials should be addressed to H. X. Tang. (e-mail: hong.tang@yale.edu).

# 1. Device fabrication and characterization

Following the procedure outlined in Ref. [15], the devices are fabricated from 110nm silicon-on-insulator substrates. Waveguides, grating couplers, and the racetrack cavity are defined by electron beam lithography in hydrogen silsesquioxane (HSQ) resist and subsequently etched in an inductively-coupled chlorine plasma reactive ion etch. For the release of the mechanical resonators a window in a photoresist mask is opened using electron beam lithography in ZEP520A resist and a subsequent short $O_2$ plasma to transfer the window into the photoresist. The mechanical beams are then undercut by wet-etching in buffered hydrofluoric acid, followed by critical point drying.

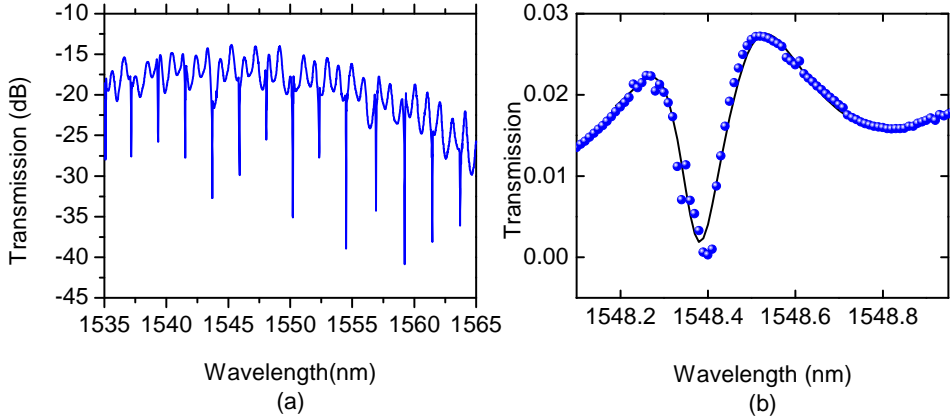

*Figure S1. (a) Measured transmission profile of the racetrack cavity. (b) Zoom of the one of the resonances with the fitted lineshape obtained from input-output theory combined with Fabry-Perot interference in the feed waveguide. The fit (solid line) yields an optical quality factor $Q_{opt} = 15k \pm 2k$*

The optical transmission of the racetrack cavity is shown in Fig. S1. It has resonances separated by a free spectral range of 2.2 nm with quality factors in excess of 10,000 and an extinction of >15 dB indicating that the cavity is close to critically coupled to the feed waveguide that carries the laser light from the input to the output coupler (Fig. 1a). The optical resonance shifts 138 pm towards longer wavelengths per mechanical resonator when they switch from buckled-up state to buckled-down state. This shift corresponds to 120 nm separation between the two buckled states.

Mechanical resonators were characterized at low optical power as well to characterize their intrinsic damping rate and resonance frequency. Figure S2 shows measured noise spectrum when nanomechanical resonators motions are dominated by thermal motion with applied Lorentzian fits. The extracted intrinsic damping rate for both resonators is $\gamma/2\pi \sim 2$ kHz.

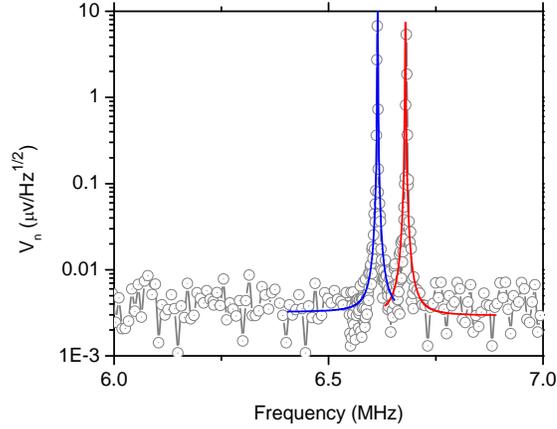

*Figure S2. Thermal noise spectrum of the nanomechanical resonators in the "up-up" state with their respective Lorentzian fits. The extracted damping rates are γ/π ~ 2 kHz for both resonators.*

## 2. Measurement setup

The measurement setup is shown schematically in Fig. 1d of the main text and it consists of two tunable semiconductor lasers (Hewlett Packard 8168F and Santec TSL210). The first one is used to probe the mechanical motion and its power is fixed during all the measurements presented here; the probe laser power in the feed waveguide is about -5 dBm. The second one is used as the pump laser to tune the backaction between the cavity and the oscillators and is tuned to a different cavity resonance than the probe laser. Also the pump laser is operated at constant power, but after being amplified by an erbium-doped fiber amplifier (EDFA) it is attenuated with a programmable optical attenuator (Hewlett Packard 8157A) to the desired level and is then filtered to suppress the unwanted amplified spontaneous emission (ASE) noise using a tunable filter with 0.1 nm bandwidth (Santec OTF900). Fiber polarization controllers are used to optimize the polarization of the pump and probe light before coupling it to the device. The latter is done using lithographically-defined grating couplers that connect to the input and output side of the feed waveguide. After the transmitted light is collected back into another single-mode optical fiber, the pump light is filtered out using an optical band pass filter (Corning 0341922001, 5 nm bandwidth), and only the probe light is detected by the photodetector (New Focus 1811). The electrical signal is then analyzed with an electrical spectrum analyzer (Hewlett Packard 4195A).

## 3. Nanomechanical resonators in the double-well potential

The buckled mechanical resonators in our devices have two stable static positions: they can either be buckled up or down. This means that the device can in principle be in four different states: "up-up", "up-down", "down-up", and "down-down", where we label the resonator with the lowest frequency in the "up-up" state as "1" and the other as "2". The thermal motion of the resonators of two of these states "up-up" and "down-up" was shown in Figs. 1(b) and (c) of the main text; we were not able to switch the second resonator to its buckled-down state through optical backaction. As has been shown previously [15], mechanical resonators switch state when they oscillate above the potential barrier. In our experimental setup, since resonator 1 has lower threshold compared to resonator 2, it starts its self-oscillation before resonator 2 does. In regime IV in Fig. 2(a) in the main text, we think that resonator 1 is above the potential barrier whereas resonator 2 is still confined within the buckled-up state potential well. As shown in Fig. 3(a), resonator 2 will eventually raise its energy above the potential barrier when the pump power is high enough. However, we were not able to access this regime experimentally.

Figure 2(a)-(g) show the evolution of the RF power spectrum of the transmitted probe laser light when the pump power increases. In regime II, there is a weak peak at the frequency difference between the oscillating resonator frequency and the resonator experiencing its thermal motion. This peak is a result of mixing between the strong oscillating signal and the thermal noise peak due to nonlinear cavity transduction at large oscillation amplitude and self-mixing in the photodetector. Also, we suspect that the latter effect gives rise to an increase in the noise floor at the low frequency end of spectrum as seen in Figs. 2(c), (e).

## 4. Analytical model

The dynamics of an optomechanical system consisting of multiple mechanical resonators with displacements $u_k$ interacting with the optical mode of a cavity with $n=|a|^2$ photons is described by the following system of coupled equations [11,12,14]

$$\frac{da}{dt} = -i\left(\Delta_0 + g_{om}\sum_k u_k\right)a - \frac{\kappa}{2}a + \frac{\kappa}{2}n_{\max}^{1/2} \tag{S1}$$

$$\ddot{u}_k = -\Omega_k^2 u_k - \Gamma_k \dot{u}_k - \frac{\hbar g_{om,k}}{m_k}n, \tag{S2}$$

where the subscript $k$ runs over the mechanical resonators; i.e. $k=1,2$ for the experiments considered here. The first equation determines the evolution of the cavity field, whereas the second is the equation of motion of the resonators. Here, $\Delta_0$ is the original detuning, i.e. the difference between the laser frequency $\omega_L$ and that of the cavity in the absence of displacements, and $\kappa$ is the optical cavity decay rate. The maximum number of photons [24] $n_{\max} = \kappa_c P_{feed}/\kappa^2 \hbar \omega_L$ is directly proportional to the power in the feed waveguide (see Fig. 1a) and to the coupling rate between the waveguide and the cavity, $\kappa_c$. The mechanical resonators have (angular) resonance frequencies $\Omega_k$, damping rates $\Gamma_k$, and masses $m_k$. To analyse this system of equations, it is convenient to switch to the complex amplitude representation for the resonators, so that their equation of motion (cf. Eq. (S2)) becomes [14, 23, 24]:

$$\dot{U}_{1,2} = \pm i\frac{\upsilon}{2}U_{1,2} - \frac{\Gamma_{1,2}}{2}U_{1,2} - i\bar{\Omega}c_{om,1,2}\langle\frac{n}{n_{max}}e^{-i\bar{\Omega}t}\rangle, \qquad (S3)$$

where $U_k = g_{om,k}\langle(u_k + \dot{u}_k/i\bar{\Omega})\exp(-i\bar{\Omega}t)\rangle = A_k\exp(i\theta_k)$ are the normalized complex amplitudes in the frame rotating at the average frequency $\bar{\Omega} = \frac{1}{2}(\Omega_1 + \Omega_2)$, $c_{om,k} = \hbar n_{max}g_{om,k}^2/m_k\bar{\Omega}^3$ are the coupling strengths, and $\upsilon$ is the frequency difference between the resonators. The latter is assumed to be much smaller than $\bar{\Omega}$ in the theoretical framework presented here so that the rotating wave approximation can be made; for the numerical simulations discussed below, this approximation is not made.

The next step is to consider the cavity response to the mechanical motion. For a single resonator the normalized amplitude-dependent cavity response function $\Phi(A) = \langle\frac{n}{n_{max}}e^{-i\bar{\Omega}t}\rangle/U$ can be calculated analytically for harmonic motion and numerically for the general case [23, 24]. The real part of $\Phi$ corresponds to reactive backaction forces (i.e. those proportional to the displacement), whereas the imaginary part describes the dissipative part (velocity proportional). Self-sustained oscillations start when the imaginary part of $\Phi(A = 0)$ exceeds $\Gamma/2\bar{\Omega}c_{om}$ and the amplitude of the resulting limit cycle $A$ is determined by the requirement $\text{Im } \Phi(A) = \Gamma/2\bar{\Omega}c_{om}$. The phase of the single OMO, however, remains undetermined.

Equation (S1) shows that when multiple resonators are present, the cavity only feels the *combined* effect of all resonators. The same reasoning as for the single OMO case leads to the conclusion that for multiple resonators coupled to the same cavity, the cavity response at any time is $\Phi = \Phi(A_+)$ where $U_+(t) = \sum_k U_k(t) = A_+(t)\exp(i\theta_+(t))$. It is important to note that the magnitude of the sum ($A_+ = |U_+|$) depends on the phase difference between the individual oscillators $\theta_2 - \theta_1$, but is independent of the global phase $\theta_+$. The equations of motion for the complex amplitudes of the two OMOs thus become:

$$\dot{U}_{1,2} = \pm i\frac{\upsilon}{2}U_{1,2} - \frac{\Gamma_{1,2}}{2}U_{1,2} - i\bar{\Omega}c_{om,k}\Phi(A_+)U_+. \qquad (S4)$$

The coupling between the resonators happens in two ways: first of all via $U_+$, and secondly via the cavity response function $\Phi(A_+)$. When the phase difference between the oscillators is a multiple of $\pi$, $U_+$ is in phase with both $U_1$ and $U_2$. In that case, the dissipative part of the backaction couples the velocity of one to that of the other and the reactive part couples their displacements. For different values of the phase difference, the displacements and velocities are all coupled to each other by the cavity backaction. Synchronized motion of the two oscillators implies that they rotate at the same rate (which could differ from $\bar{\Omega}$ by a frequency offset $\epsilon$). This means that then $U_{1,2} = Y_{1,2}\exp(i\epsilon t)$ must be a solution to Eq. (S4) for time-independent $Y_{1,2}$. After inserting this solution into Eq. (S4), combing terms, and adding, the equation that determines the combined amplitudes of the limit cycles is obtained:

$$\Phi(A_+) = \frac{1}{\overline{\Omega}} \frac{\left(\frac{v}{2}\right)^2 - \epsilon^2 + i\epsilon\bar{\gamma} + \frac{i\,\delta\gamma v}{4} + \left(\frac{\bar{\gamma}}{2}\right)^2 - \left(\frac{\delta\gamma}{4}\right)^2}{2\epsilon\bar{c}_{om} + v\frac{\delta c_{om}}{2} + i\frac{\delta c_{om}}{2}\frac{\delta\gamma}{2} - i\bar{\gamma}\bar{c}_{om}}, \tag{S5}$$

where $\bar{c}_{om}$ ($\delta c_{om}$) and $\bar{\gamma}$ ($\delta\gamma$) are the average (difference in) coupling strengths and damping rates respectively. Note that the asymmetries ($\delta c_{om}$, $\delta\gamma$, and $v$) either appear in pairs or quadratically in Eq. (S5), and that for two identical oscillators the Eq. (S5) reduces to that of a single oscillator with twice the original coupling strength: $\Phi(A_+) = (\frac{i\gamma}{2} - \epsilon)/2\overline{\Omega}\bar{c}_{om}$ as expected for in-phase motion.

### 4.1. Thermal force noise and impulse response

Thermal force noise $F_{\text{th},k}$ acts on the two resonators and appears on the right hand side of Eq. (S2). In the absence of oscillations this results in a small Brownian motion with root-mean-square amplitude $\sqrt{k_B T/m_k \Omega_k^2}$, where $k_B$ is Boltzmann's constant. For our device this amplitude is ~ 100 pm at a temperature T = 300 K. Instead when the resonators are oscillating with large displacements, the force noise causes small amplitude and phase fluctuations of the carrier. The effect of the force noise can be understood using the impulse response. Any linear system is characterized by the impulse response function $g(t;t')$ which is the displacement $u(t)$ generated by a force $\delta(t-t')$. Here $\delta(t-t')$ is the Dirac delta function which corresponds to a kick of the resonators at time $t = t'$. Integration of the acceleration over an infinitesimal interval around $t = t'$ shows that the immediate result of the kick is a step in the velocity. By calculating the Fourier transforms for the responses to a kick at different (random) times, the same spectrum is obtained as for white force noise[1]. In the numerical simulations the effect of force noise can thus be accounted for via the impulse response without having to integrate the full stochastic differential equations.

---

[1] In this single-kick case the relative height of the carrier and sidebands in the Fourier transform depend on the length of the timetrace as the kick-induced transient dies out whereas the oscillations keep going. For true thermal force noise that is not the case as the kick happens continuously. In principle this can be fixed by generating the timeseries for a large number of kicks (at random times and with a low-frequency power-spectral density corresponding to the thermal force noise) and subsequent ensemble averaging the spectra over many different realizations as shown in Sec. 5.1. This, however, requires very long timeseries that are about the number of kicks times larger than the timeseries for a single kick. Hence for the power dependence in Fig. 3 we simulated the response to a single kick which still gives the right spectrum, although comparing the magnitude of the sidebands to the carrier is now no longer possible.

# 5. Numerical simulations

Computer simulations of the coupled cavity-resonators system are done by numerically integrating Eqs. (S1) and (S2) using a fourth order Runge-Kutta algorithm with adaptive step size implemented in C++. No assumptions are made on the smallness of any parameter, including the frequency difference $\upsilon$. For the simulation of the double well potential (but not for the injection locking simulations), the $-\Omega_k^2 u_k$ term in Eq. (S2) was replaced by $+\Omega_k^2(u_k - \alpha_k u_k^3)$ and the detuning was defined with respect to the right potential minima at $u_k = \alpha_k^{-1/2}$. The simulated time is taken long enough to ensure that transient effects are negligible and that stable SSOs result. Then the two resonators are kicked once (i.e. their velocity is suddenly slightly modified while keeping the displacements and cavity field the same) to simulate the effect of thermal noise. Fourier transforms are calculated in Matlab with a flattop window function that provides a steep roll-off without too much computational costs. This results in a background that is more than 100 dB below the carrier as shown in Fig. S3(a). By repeating this procedure a number of times (10 for the simulations shown here) with pseudo-random kick times and subsequent averaging of the magnitude of the spectrum, the smooth curves of Fig. 3(c) and S3(a) are obtained. Note again that without the kicks (or, equivalently, with zero kick strength), sidebands are never observed in the dynamics of the steady state SSOs. Figure S3(b) shows the magnitude of the carrier and sidebands extracted at different kick strengths. At low kick strengths $k$ the sidebands do not appear and the magnitude at their frequency is equal to the background due to the carrier $y_0$. Note that without the flattop window this background would be much higher (~30 dBc) and the sidebands would not be resolvable. When the sidebands appear at $k \sim k_0$ their magnitude scales linearly with $k$. When the background and the sidebands do not interfere, the total magnitude should be given by $y = ((k/k_0)^2 + y_0^2)^{1/2}$ and the fitted black line shows that this is indeed the case.

For the simulations in Fig. 3 a kick strength of $k = 0.07$ has been used. Finally, we have simulated the dynamics of the system in the presence of noise for a single set of parameters. Here, instead of a kick the true stochastic thermal force is included. Figure S4 shows that this broadens the carrier due to the induced phase diffusion. Also, as expected the same sidebands as in the kicked oscillator case (Fig. S3(a)) show up and their magnitude of about -40 dBc. This is only a factor of 3 below those seen in the experiment (-35 dBc, see main text). The full stochastic simulations thus allow for a more quantitative comparison of the magnitude of the sideband vs. the carrier, but are much more computationally expensive as explained above.

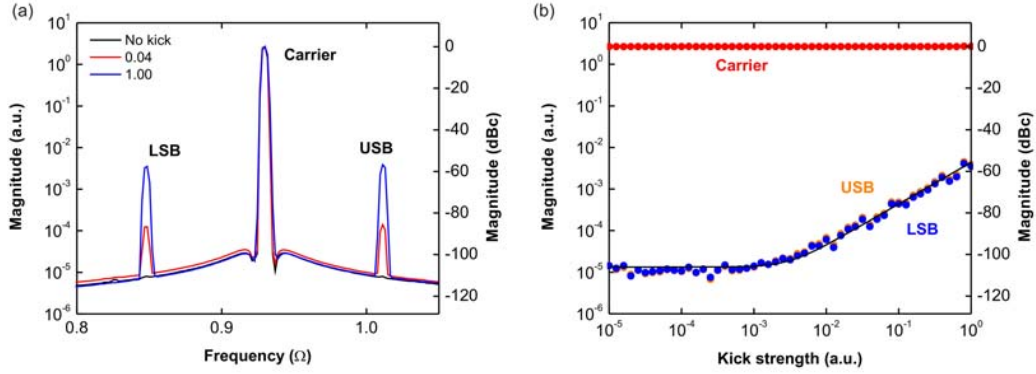

*Figure S3. (a) Fourier spectra of the simulated cavity occupation. First the system let to reach steady oscillations and is then kicked with varying strength. Without a kick (black), only the carrier at $0.93\bar{\Omega}$ is visible. For finite kick strength (red, blue) lower and upper sidebands (LSBs and USBs resp.) appear symmetrically around the carrier. (b) The magnitude of the peaks versus the kick strength. For very small kick strengths the sidebands are masked by the spectral leakage of the carrier, but for kick strengths $>> 10^{-3}$ the sideband magnitude is proportional to the kick strength (black line). The parameters in these simulations are the same as in Fig 3c of the main text and $n_{max} = 50$.*

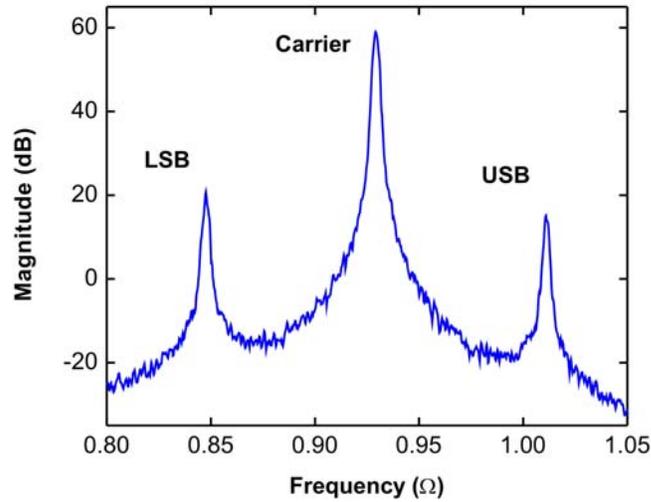

*Figure S4. Averaged power spectral density of simulated timetraces with noise. The parameters are the same as for Fig. S3. The force noise strength was chosen such that the ratio of the intrinsic (i.e. without the cavity) thermal motion over the separation between the two wells was $10^{-3}$ which is the same magnitude as in the experiment.*

## 5.1. Synchronization in the time domain

To further illustrate the synchronous dynamics of our oscillators, the time evolution of the region in Fig. 3a with a strong carrier tone is studied in more detail. Figure S5 shows the evolution of the complex amplitudes $U_{1,2}$ at $n_{max}$ = 956. For this power the two oscillators are both above the potential barrier and their motion amplitude (i.e. the length of the sticks) is similar. Also, the angle of the complex amplitude rotates in time, indicating a difference between the oscillation frequency and the reference frame rotating at $\bar{\Omega}$, due to a finite value of $\epsilon$ (see Eq. (S5)). However, note that the (small) angle between the two amplitudes remains constant in time, indicating that the two oscillators are indeed synchronized with a small constant phase difference as predicted by Eq. (S4).

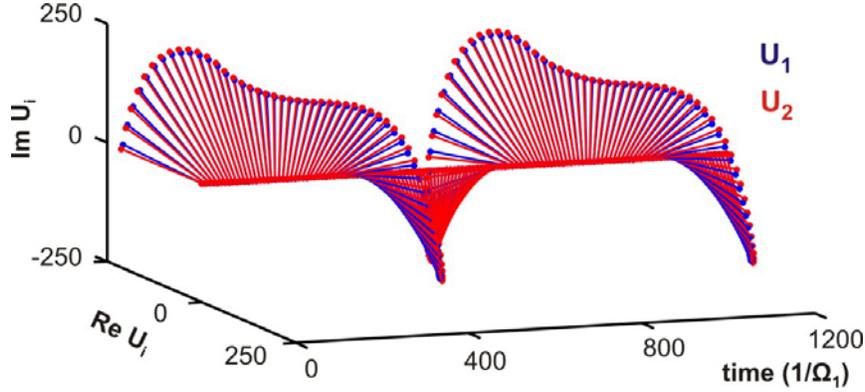

*Figure S5. Complex amplitudes of the two oscillators obtained from a simulated timetrace corresponding to $n_{max}$ = 956 in Fig 3c of the main text.*

This can be further illustrated by extracting the phase difference between them at times that are spaced by exactly one oscillation period. In such a corotating frame, the effect of a finite $\epsilon$ is compensated for, and the stroboscopic sampling prevents artefacts of the nonlinear potential. Figure S5 shows the phase evolution of the phase difference for the data of Fig. 3 for $n_{max}$ = 50. Without a kick the phase difference remains constant at about 16.1 mrad (red dashed line). The kick at $t$ = 0 appears as a jump in the phase difference, which overshoots a bit, and then decays back to the original value of 16.1 mrad in about 500 periods. The fact that the phase difference returns back to exactly the same value thus proves that the phase difference between the two synchronized oscillators is fully coherent. On a much smaller timescale, oscillations in the phase are visible and these will appear as (frequency-modulation) sidebands around the carrier. Their period is 4.78 oscillator periods, which corresponds to a frequency of 0.21 $\bar{\Omega}$. Comparing with Figure 3(c) shows that this frequency indeed corresponds to the position of the sidebands relative to the carrier.

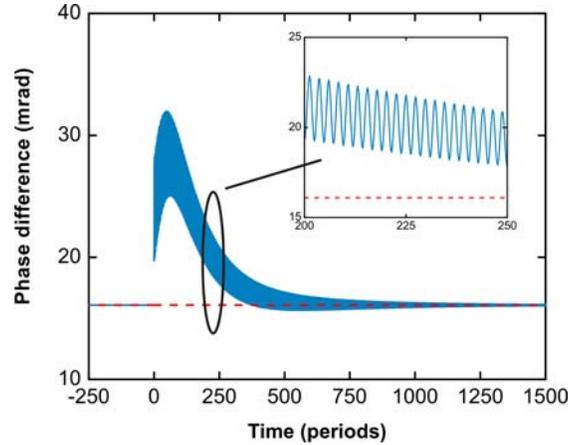

*Figure S6. Evolution of the phase difference between the two oscillators with (blue) and without (red dashed) a kick at t=0. A zoom of the same data is shown in the inset. The kick strength for the blue curve is k=0.1 and $n_{max}$ = 50 for both curves.*

## 5.2. Injection locking

The injection locking experiment is emulated with the same code, but now the backaction to the second resonator is switched off ($c_{om,2}$ = 0) and it is driven with a strong coherent force at the modulation frequency. Similar to the experimental situation where the input laser power is modulated, the second resonator motion modulates the cavity occupation. Also here, sidebands are only observed when kicking oscillator 1. Also, note that these simulations are for harmonic oscillators. The observed sidebands are thus not due to mechanical nonlinearities.

Figure S7 shows the results for the injection locking simulations. At low modulation depth the oscillator (red line at $\Omega$ = 1.26 $\Omega_1$) does not lock to the modulation (the orange line at $\Omega$ = 1.05 $\Omega_1$) and, since both the oscillator and modulation amplitudes are large. Again their mixing products show up as a multitude of sidebands of the two strong tones. When the modulation is strong enough the oscillator locks to it and only a single strong tone is visible. Just as in the experiment, the sideband frequency offset increases with increasing modulation depth.

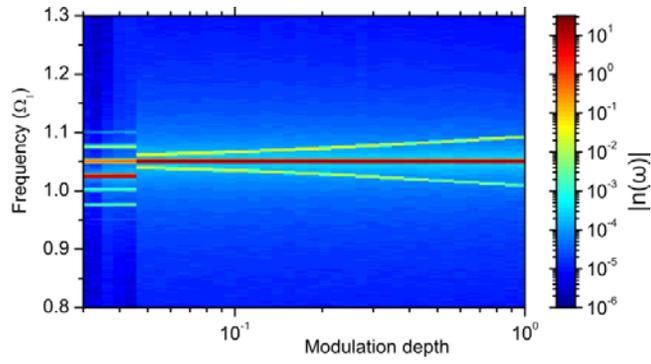

*Figure S7. Colormap of the Fourier transform of simulated timetraces of the cavity occupation n(t) for a modulated power in the cavity. The oscillation carrier appears in red. For small modulation depth (m <0.046) the oscillator does not lock to the modulation at $1.05\Omega_1$ (orange), but for large drive (m >0.046) it locks to the modulation tone and sidebands (yellow/cyan) emerge that increase in offset frequency with increasing modulation strength.*